# Proximity driven commensurate pinning in $YBa_2Cu_3O_7$ through all-oxide magnetic nanostructures

M. Rocci,[†,&] J. Azpeitia,[⊥,&] J. Trastoy,[‖] A. Perez-Muñoz,[†,&] M. Cabero,[†,&] R. F. Luccas,[⊥,&] C. Munuera,[⊥,&] F. J. Mompean,[⊥,&] M. Garcia-Hernandez,[⊥,&] K. Bouzehouane,[‖] Z. Sefrioui,[†,&] C. Leon,[†,&] A. Rivera-Calzada,[†,&] J.E. Villegas,[‖] and J. Santamaria[†,&*]

[†]GFMC, Dpto. Fisica Aplicada III, Univ. Complutense Madrid, 28040 Madrid, Spain,

[⊥]Instituto de Ciencia de Materiales de Madrid, 28049 Madrid, Spain,

[&] Unidad Asociada Laboratorio de Heteroestructuras con Aplicación en Espintrónica" UCM-CSIC, 28049 Madrid, Spain,

[‖]Unité Mixte de Physique, CNRS, Thales, Univ. Paris-Sud, Université Paris Saclay, 91767 Palaiseau, France.

The design of artificial vortex pinning landscapes is a major goal towards large scale applications of cuprate superconductors. While disordered nanometric inclusions have shown to modify their vortex phase diagram and to produce enhancements of the critical current[1,2], the effect of ordered oxide nanostructures remains essentially unexplored. This is due to the very small nanostructure size imposed by the short coherence length, and to the technological difficulties in the nanofabrication process. Yet, the novel phenomena occurring at oxide interfaces open a wide spectrum of technological opportunities to interplay with the superconductivity in cuprates. Here we show that the unusual long range suppression of the superconductivity occurring at the interface between manganites and cuprates affects vortex nucleation and provides a novel vortex pinning mechanism. In particular, we show evidence of commensurate pinning in YBCO films with ordered arrays of LCMO ferromagnetic nanodots. Vortex pinning results from the proximity induced reduction of the condensation energy at the vicinity of the magnetic nanodots, and yields an enhanced friction between the nanodot array and the moving vortex lattice in the liquid phase. This result shows that all-oxide ordered nanostructures constitute a powerful, new route for the artificial manipulation of vortex matter in cuprates.

The discovery of the metallic state at the interface between LAO and STO[3], two band insulators, has driven a strong effort to understand its origin and to functionalize its exotic properties into devices[4]. The technological opportunities may be dramatically expanded if instead of relatively simple band insulators more complex (correlated) transition metal oxides get in touch[5,6]. Many of these correlated oxides share a common perovskite structure which allows combining them in highly perfect epitaxial interfaces to bring very different ground states into direct contact. This is the case of the hetero-epitaxial interfaces between ferromagnetic manganites and high Tc superconductor cuprates, which host interesting forms of interplay between magnetism and superconductivity, resulting from induced magnetism in the cuprate[7–10]. While the induced spin polarization in the cuprate is probably at the origin of the exotic proximity phenomena observed at these interfaces[11,12], such as the equal spin Andreev reflection by which triplet pairing emerges at these interfaces[13], the long range suppression of superconductivity[14–16] can affect vortex nucleation and be the source of novel pinning phenomena when the superconductor is driven to the mixed state by the application of an external magnetic field.

In this paper, we examine the use of the long range suppression of superconductivity occurring at the interface between cuprates and manganites to design artificial pinning landscapes. In particular we show periodic pinning by an array of manganite magnetic nanodots in YBCO layers of nanometric thickness. Periodic pinning by (sub)micrometric antidots (holes) was shown in YBCO[17] and BSCCO[18,19], and more recently by irradiating with high energy oxygen ions through a PPMA mask with a hole (30-80 nm diameter) array defined by electron-beam lithography[20–22]. However, periodic pinning by an array of magnetic nanodots in transition metal oxide ferromagnet/cuprate superconductor hybrids has to our knowledge never been shown. This is probably due to technological problems in sample fabrication, mainly related to the sensitivity of the structure of oxides to ion etching processes and to their chemical similarity, which makes selective etching processes difficult to identify. Moreover, while in the case of transition metal systems periodic pinning is often governed by the dipolar coupling of the vortex moment to the stray fields created by the magnetic dots[23], in oxides peculiarities of the domain structure and the long range interplay between both (ferromagnetic and superconductor) long range orders make the system an attractive scenario to explore. Apart of the basic interest, the design of artificial pinning sites is highly interesting for practical applications of the high Tc superconductivity.

An array of circular shaped (100 nm diameter) magnetic dots was defined in 25 nm thick manganite (LCMO and LSMO) layers by electron beam lithography (Figure 1a and b). The inter-dot distances are 300 nm and the array is defined over an area of on 100 x 100 micron squared. A 20-50 nm thick YBCO layer was deposited ex situ. Optical lithography was used to define an 8-probe bridge for electrical transport measurements (inset to Figure 1c).

The Tc of the YBCO films on the nanodot arrays (measured from the resistive transition) was not significantly depressed as compared with the single layer one (Figure 1c). The magnetoresistance curves – measured at temperatures close to Tc with the magnetic field applied perpendicular to the layers surface, see Figure 2 – show produced pronounced dips at magnetic field values $B_\Phi$ which yield a vortex density $\frac{1}{a_{vortex}^2} = \frac{B_\Phi}{\Phi_0}$ matching the dots array areal density $\frac{1}{a_{dots}^2} = \frac{B_\Phi}{\Phi_0}$, where $\Phi_0$ is the flux quantum. It is well known that, at this fields, geometric coincidence between the vortex-lattice and the pinning array results in enhanced pinning, leading to reduced resistance and increased critical currents[24–26]. For fixed current densities, increasing temperature results in a reduction of the resistance dips depth, as shown in Figure 2a a for YBCO grown on a ferromagnetic $La_{0.7}Ca_{0.3}MnO_3$ dots array. On the other hand, for a fixed temperature, increasing the current density also results in a reduction of the resistance dips depth (see Figure 2b). To compare the effect of pinning it may be instructive to compare the resistance at matching fields and out-of-matching. The resistance is associated to the appearance of an electric field $\boldsymbol{E} = \boldsymbol{v} \times \boldsymbol{B}$ when, for an applied magnetic field $\boldsymbol{B}$, vortices move with velocity $\boldsymbol{v}$ under the action of a Lorentz force per unit (vortex) length $\boldsymbol{J} \times \boldsymbol{\Phi_0}$. Here $\boldsymbol{J}$ is the current density. The resistance is thus a direct measure of vortex-velocity, which can be quantified at matching and out-of-matching fields. Thereafter, we denote "out-of-matching field" to the field at which a resistance maximum is observed just below the matching field (see arrows in the inset to Figure 3). The main panel of Figure 3 displays the vortex velocity as a function of temperature for fixed current levels (Lorentz forces), both at matching (hollow symbol) and out-of-matching fields (solid symbols). For each current (see labels) and temperature, the difference in velocity between hollow and solid symbols (see double-head arrows) measures the enhancement of the interaction between the vortex-lattice and the pining array at matching condition. In other words, comparison of hollow/solid symbol measures by how much the vortex-lattice is slowed down due to the enhanced interaction between vortices and the pining array at the matching condition. From the inspection of that Figure, it is clear that the matching effects are relatively more pronounced the lower the temperature and the lower the current density, which indicates that the periodic pinning effects are maximized in the vortex liquid phase. Notice also that the matching effects are stronger the smaller vortex velocity, and that for a given vortex velocity range, the velocity decrease at matching is similar regardless of the temperature and applied current (Lorentz force). It is clear, then, that the periodic pinning effect can be viewed as a friction of the vortex lattice with the underlying dot array, which is maximized at the matching condition.

We can rule out Ca diffusion and other forms of intermixing as the origin of the observed pinning phenomenon since scanning transmission electron microscopy (STEM) and electron energy loss spectroscopy (EELS) with atomic column resolution have shown that the interfaces are both structurally and chemically abrupt at atomic scale.[27]

To investigate the role of the manganite magnetic ground state in the pinning phenomenon we prepared samples with the same array geometry and YBCO thickness (20 nm) as those discussed above, but with $La_{0.3}Ca_{0.7}MnO_3$ dots which is known to be insulating and antiferromagnetic (AF) below 250 K (instead of half-metallic and ferromagnetic). In these samples, the Tc of the YBCO was Tc=85K. While the signature of periodic pinning was also found in these AF-nanodot samples this was in all cases much weaker than with the ferromagnetic dots. This is shown in Figure 2c. Note that in Figure 2 c the minima are shallower than Figure 2b, and the second order matching is hardly visible. Nevertheless, from the observation of matching effects in some AF-insulating nanodots based samples, it is clear then that the YBCO layer corrugation plays a role in the pinning effect, probably as the result of the reduction of the vortex length in the YBCO grown on top of the dots. AFM observations of the surface topography after the growth of the YBCO (not shown) showed the height modulation due to the dots underneath but with a reduced amplitude (6-10 nm instead of the 24 nm corresponding to the dot thickness). This evidences that the growth of the YBCO smoothens the distribution of heights probably as the result of lateral growth from the lateral surface of the dots. As a result there is a distribution of vortex lengths (short vortices on the dots and long vortices out of the dots). Since vortex energy scales with vortex length, this could have an effect on pinning as previously proposed by Daldini and collaborators[28]. However, in view of the stronger pinning effect consistently found in the samples with ferromagnetic dots, we conclude that magnetism of the dots plays a role in the periodic suppression of dissipation.

To learn about the magnetism state of the ferromagnetic manganite nanodots, magnetic force microscopy experiments were conducted on a patterned dot array (prior to the YBCO growth). Figure 4a and b display MFM images taken at 125 K with a 5000 Oe magnetic field applied perpendicular to the plane (Figure 4a) and in the plane applied along the diagonal direction of the dots array (Figure 4b). A clear magnetic contrast evidences a large saturation magnetization of the dots for both field geometries. Reducing magnetic field to the range (0 to 500 Oe) in which magnetotransport measurements are conducted results in a loss of magnetic contrast. This indicates a low remanence magnetic state (see Figure 4c), consistent with the nucleation of small domains. This shows that the stray magnetic fields from the dots, which are directly probed by MFM, must be small and, thus, that the dipolar coupling between the flux quanta (vortices) and the nanodot stray fields can be discarded as the origin of the observed matching effects. Moreover, the magneto-transport curves and the strength of the matching effects were independent on the magnetic history. As can be seen in Figure 4d, no change was observed in the depth of the of the R(H) dips after field cooling in H=5000 Oe applied out of plane and after demagnetization (magnetic field cycles with decreasing field amplitude and alternating polarity, from 600 Oe to 0 Oe in 10 Oe steps). Also, no change in R(H) was observed in measurement performed after magnetic saturation in H=5000 Oe, neither for in-plane nor for out-of-plane applied fields. Again, this result indicates that the dots are in a multi-domain state within the low magnetic fields range in which the resistance minima are observed and, as a consequence, that the origin of the enhanced pinning must be other than dipolar coupling of the vortices to the stray fields from the dots.

We propose that the long range suppression of the superconducting order parameter reported at these ferromagnet superconductor interfaces is at the origin of the observed pinning phenomenon[14,15]. At cuprate manganite interfaces the Mn-O-Cu bond resulting from the hybridization between reconstructed 3d $3z^2$-$r^2$ orbitals constitutes an antiferromagnetic (AF) superexchange path. The AF interaction is transmitted to the 3d $x^2$-$y^2$ Cu orbitals by Hund coupling interaction[29]. As a result of the effective Mn-O-Cu AF superexchange, a form of magnetism is induced in the cuprate at the interface controlled by Mn interfacial moments, as it has been shown by x ray magnetic circular dichroism, XMCD. I.e., the Cu magnetic state is ferromagnetic-like with remanence and coercivity imposed by the interfacial Mn magnetic moments. The spin polarization induced in the cuprate conduction band[30,31] underlies the long range suppression of the superconductivity at these interface occurring with a length scale (10 nm) orders of magnitude larger than the c-axis superconducting coherence length (0.1 – 0.3nm)[14,16,32]. The long length scale proximity effect with 10 nm length scale observed in trilayer and superlattice samples involving two F/S interfaces indicates that in fact superconductivity is depressed over 5 nm from the (single) interface as it has been theoretically described[14–16]. It is precisely this length scale (around the ferromagnetic dots) over which the pinning is promoted due to the reduced cost of condensation energy.

To get further insight into the pinning mechanism we prepared samples with the same dots array and with oxygen depleted YBCO. The effect of oxygen outtake from YCBO is increasing the vortex anisotropy. This reduces the intensity of the native pinning strength and, more importantly, it provides a measure of the homogeneity of the pinning mechanism along the vortex line. The loss of correlation along the vortex line upon increasing anisotropy causes a vortex dimensional crossover, from 3D for fully oxygenated samples to quasi 2D for x=6.55 (when vortex correlation length is smaller than sample thickness) to pure 2D for x=6.35, close to the superconductor to insulator transition[33]. Magneto-resistance curves for samples with different reduced oxygen content are shown in Figure 5a and b. While we observe a small increase of the pinning strength for small oxygen outtakes corresponding to x=6.6 (see Figure 5a), further reducing the oxygen content decreases the intensity of the resistance dips, which essentially disappear for a sample with Tc=37 K corresponding to an oxygen content of 6.45 (see Figure 5b), in the quasi 2D vortex state. Wide regions of the vortex phase diagram were explored by changing the current density and temperature to make sure that the commensurate pinning did not show up.

To quantitatively compare the intensity of the matching effects in the different samples, it is useful to plot the vortex velocity as a function of the current density (Lorentz force). This is done in Figure 5c, with hollow symbols for the matching field and solid one for out-of-matching measurements. Let us remember at this point that we learnt from Figure 3 that comparisons between samples can be done for temperature/current combinations that yield vortex velocities within the same range. For each sample, the strength of the matching effects can be quantified along the vertical (velocity) axis, in terms of the vortex velocity decrease at the matching (hollow) relative to the out-of-matching (solid) field, at constant Lorentz force. By using the velocity variation criterion, it becomes clear that in the case of antiferromagnetic LCMO dots (red symbols) the matching effects are less intense than for ferromagnetic dots (blue solid symbols), which indicates the ferromagnetism plays a role in the pinning mechanism. Also notice that pinning strength increases for the slightly deoxygenated sample (x=6.6) (solid and open squares , ■,□) as compared to the fully oxygenated (x=6.95) sample (solid and open circles ●,○), probably as a result of the comparatively less intense native pinning in the more anisotropic x=6.6 sample. The pinning strength drops dramatically for the more deoxygenated sample (x=6.45). We interpret this as the result of the loss of correlation along the vortex line at the dimensional transition taking place in oxygen depleted samples. I.e, in the quasi 2D vortex state there is a loss of vortex line tension when correlation length becomes shorter than sample thickness. The strong decrease of the pinning strength observed in the quasi 2D vortex state indicates that pinning is inhomogeneous along the vortex line: the top of the vortex line becomes free to dissipate even when its lower part is pinned. This result nicely supports the interpretation that the pinning mechanism being driven by the magnetic proximity suppression of the superconductivity at the ferromagnetic dots (the shaded area in the sketch displayed as an inset to Figure 5).

In summary, we have shown commensurate pinning in YBCO by an ordered array of manganite dots. By comparing the effect of ferromagnetic and antiferromagnetic dot arrays, we have shown that the pinning phenomenon has magnetic origin. The geometric corrugation effect caused by the thickness modulation of the superconductor as it grows on the manganite nanostructures has also a pinning effect although it is comparatively less intense than the magnetic contribution. The small remanence of the dots shown by the MFM observations rules out the dipolar coupling of the vortex current field to the stray fields of the vortices as the dominant pinning contribution. We conclude that pinning originates at the proximity suppression of the superconducting order parameter at the interface with the manganite, which as it is well-known is due to the build-up of a spin polarization caused by the Mn-O-Cu superexchange interaction.

**Methods: Device Fabrication and Experimental Setup.**

Samples were grown on (001) SrTiO$_3$ single crystals by a high pressure (3.4 mbar) pure oxygen sputtering technique at high temperatures (900 ºC) from stoichiometric target which produces good epitaxial growth of cuprates and highly ordered manganite cuprate interfaces[34]. Oxygen content of the YBCO was adjusted following a stability line of the pressure-temperature phase diagram[14,33]. We produced samples with oxygen contents of x=6.6 (Tc=47 K) and x=6.45 (Tc=35 K). It is important to notice that oxygen outtakes from the YBCO did not affect the manganite grown underneath[35]. Dot arrays were fabricated by electron beam lithography (Raith 50) followed by wet etching. Atomic/magnetic force microscopy (AFM/MFM) measurements were performed in a commercial Low-Temperature SPM equipment from Nanomagnetics Instruments, working in the 300K-1.8K temperature range. The microscope is compatible with a home-made three axis superconducting coil system, which allows performing experiments under applied magnetic fields of up to 5T along Z-axis and 1.2 T for X-Y plane[36]. Simultaneous magnetic and topographic images shown in this work were obtained at 125K, using the retrace mode at a lift distance of ~80nm, employing commercial tips from Nanosensors (PPP-MFMR). Prior to MFM measurements, the tips were magnetized with an external field (500 Oe) in their axial direction. Magnetoresistance measurements were performed in a closed cycle cryostat equipped with a 4000 Oe electromagnet and in a Quantum Design 9T PPMS system equipped with a horizontal rotator.

ACKNOWLEDGMENTS

JS acknowledges CNRS for a stay at Unite Mixte de Physique CNRS Thales. Work at UCM supported by grants MAT2014-52405-C02-01 and Consolider Ingenio 2010 - CSD2009-00013 (Imagine), by CAM through grant CAM S2013/MIT-2740. Work done at ICMM supported by grants MAT2011-27470-C02-02, MAT2014-52405-C02-02 and CSD2009-00013. J. E. V. thanks European Rerearch Council for financial support with grant No. 647100 "SUSPINTRONICS".

FIGURE CAPTIONS

**Figure 1**. (a) Square array (300 x 300 nm) of manganite dots fabricated by electron beam lithography. (b) Enlarged view of a single dot 100 nm diameter made of 20 nm thick LSMO. (c) Resistive transition of YBCO layers (24 nm in red and 50 nm in blue) grown on top of the dots and patterned as show in the inset for transport measurements.

**Figure 2**. Magnetoresistance of YBCO (50 nm) grown on LCMO magnetic dots (20 nm) in magnetic fields ranging between -1000 and 1000 Oe applied perpendicular to the layer surface. (a) Ferromagnetic $La_{0.7}Ca_{0.3}MnO_3$ dots, at fixed current of 10 µA and changing temperature (T=86.6 K (blue), T= 86.7 K (purple) and T=86.9 K (red)). (b) Ferromagnetic $La_{0.7}Ca_{0.3}MnO_3$ dots, at fixed temperature and changing current (I=10 µA (blue), I= 100 µA (purple) and I= 500 µA (red). (c) Antiferromagnetic $La_{0.3}Ca_{0.7}MnO_3$ dots, at fixed temperature (T= 84 K) and changing current (I=10 µA (blue), I= 100 µA (purple) and I= 500 µA (red)).

**Figure 3**. Vortex velocity as a function of temperature computed at matching fields (open symbols) and out of matching (solid symbols) as indicated in the inset. Different colors correspond to different current levels (red (10 µA), black (100 µA), blue (1mA)). Notice that the velocity reduction at commensurability is similar for similar vortex velocities unless too far into the nonlinear regime.

**Figure 4**. Magnetic Force Microscopy Images of an LSMO dots array at 125 K at magnetic fields of 5000 Oe applied perpendicular (a) and parallel (b) to the sample surface and in remanence (c). All MFM images share a common scale. Total z-scale from 5 Deg to -5 Deg. The baseline offset was independently adjusted to maximize contrast. (d) Magnetoresistance of YBCO deposited on a similar dots array at 200 µA and 88.5 K after zero field cooling (black) and after demagnetizing the sample (red).

**Figure 5**. (a) Magnetoresistance of a deoxygenated YBCO (x=6.6) film grown on top of 100 nm LSMO dots array (300 x 300 nm) at 47 K. Currents are 10, 20, 50, 100, 200, 500,1000 µA from bottom to top. (b) Magnetoresistance of a deoxygenated YBCO (x=6.45) film grown on top of 100 nm LSMO dots array (300 x 300 nm) at 35 K. Color code is the same in (a) and (b). (c) Vortex velocity versus Lorentz force per unit length for different samples of this study in matching field (open symbols) and out of matching (solid symbols): LCMO AF dots (Red circles), LCMO FM dots (Dark blue circles), LSMO FM dots (Green up triangles and green circles). LCMO FM dots for comparison with LSMO (stars). LSMO FM dots with deoxygenated x=6.6 YBCO (Blue squares). LSMO FM dots with deoxygenated x=6.45 YBCO (Magenta Circles). Inset: sketch illustrating pinning by magnetic proximity suppression of superconductivity.

ABBREVIATIONS

AF, antiferromagnetic; FM, ferromagnetic; YBCO, $YBa_2Cu_3O_7$; LCMO, $La_{0.7}Ca_{0.3}MnO_3$; LSMO, $La_{0.7}Sr_{0.3}MnO_3$.

Figure 1

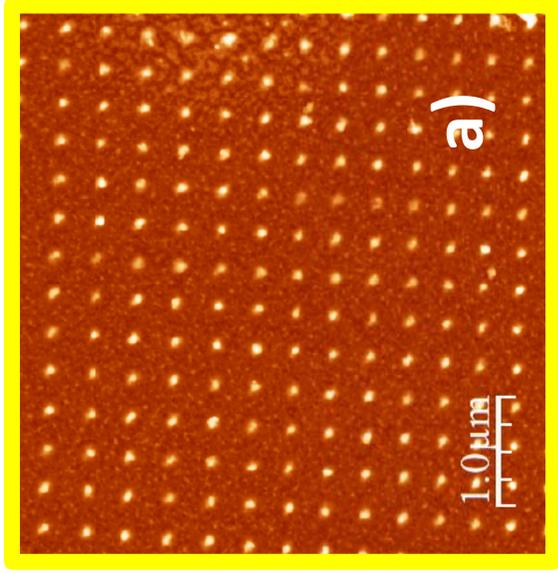

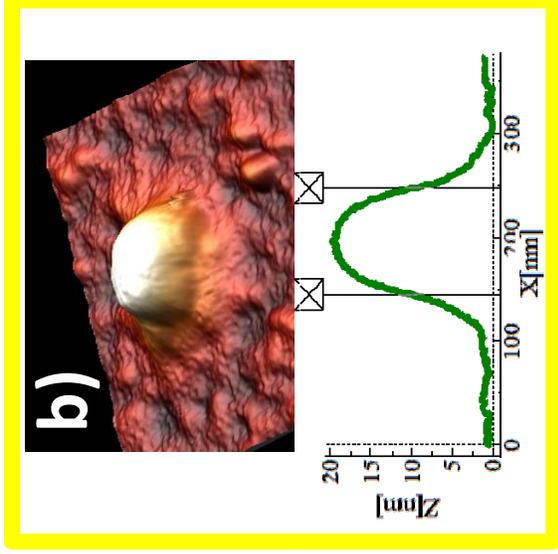

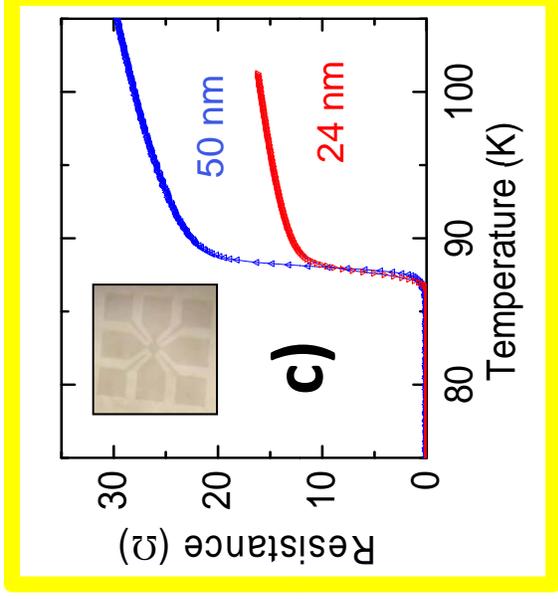

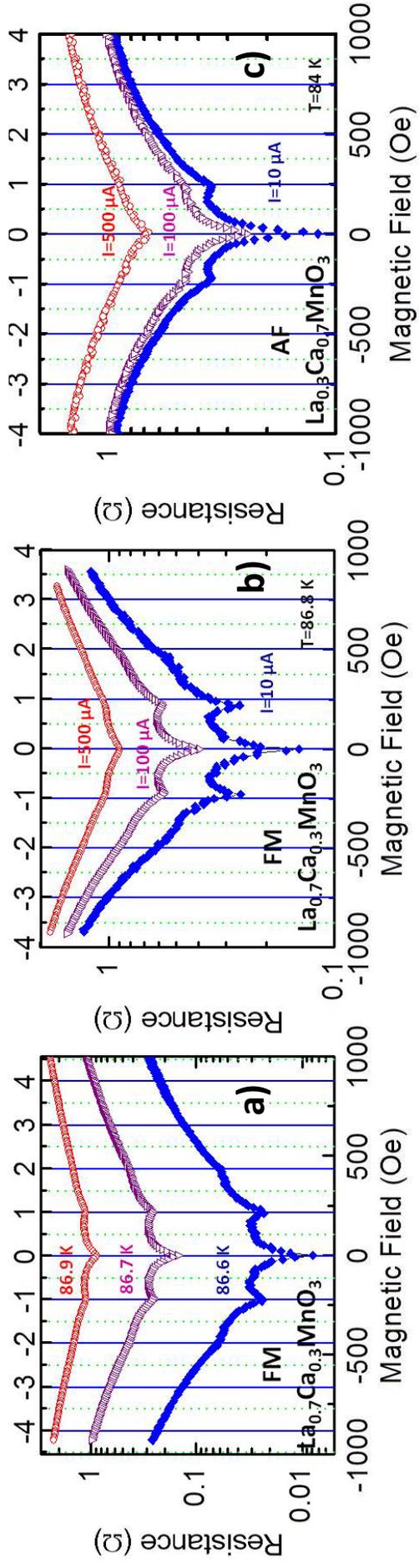

Figure 2

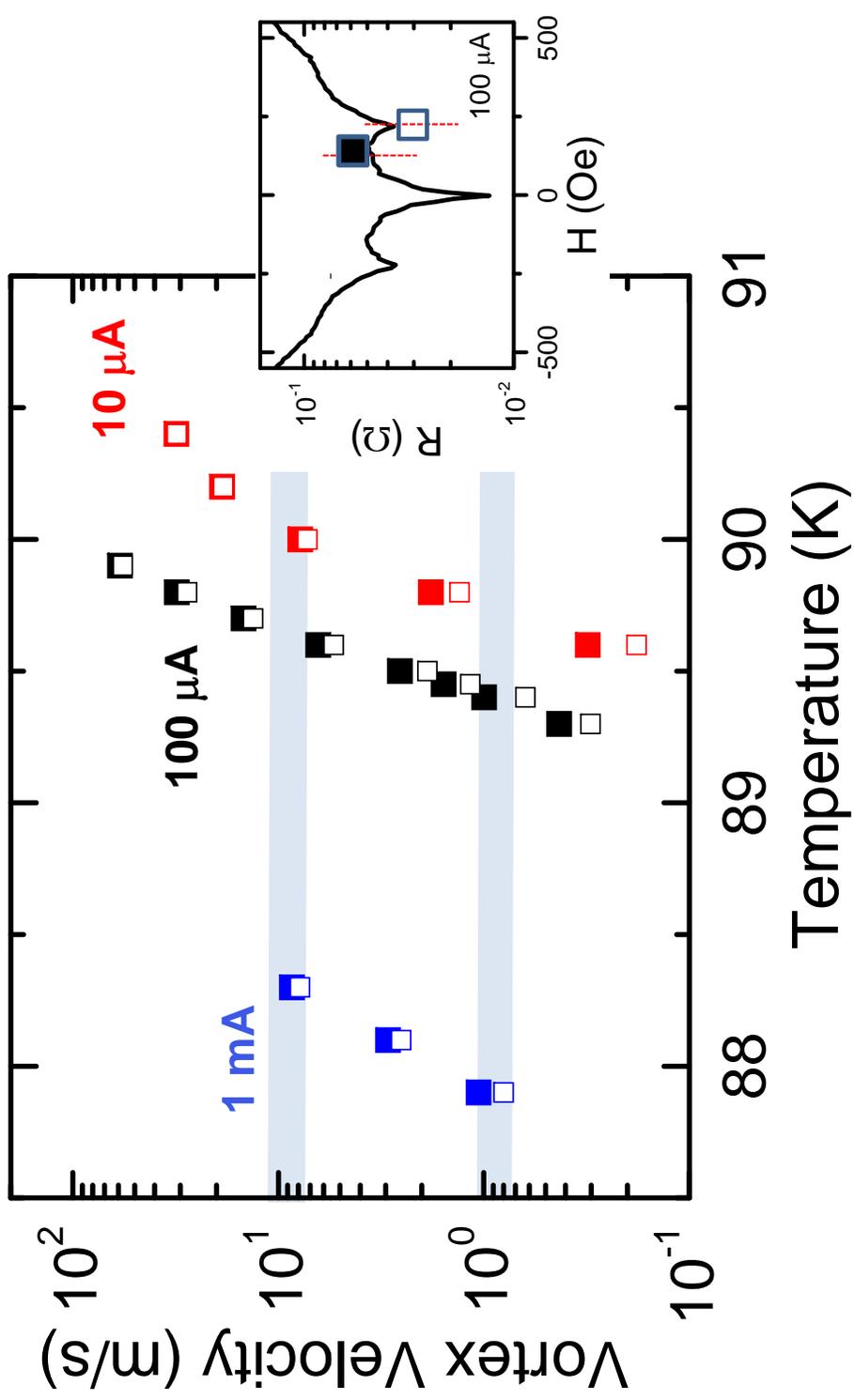

Figure 3

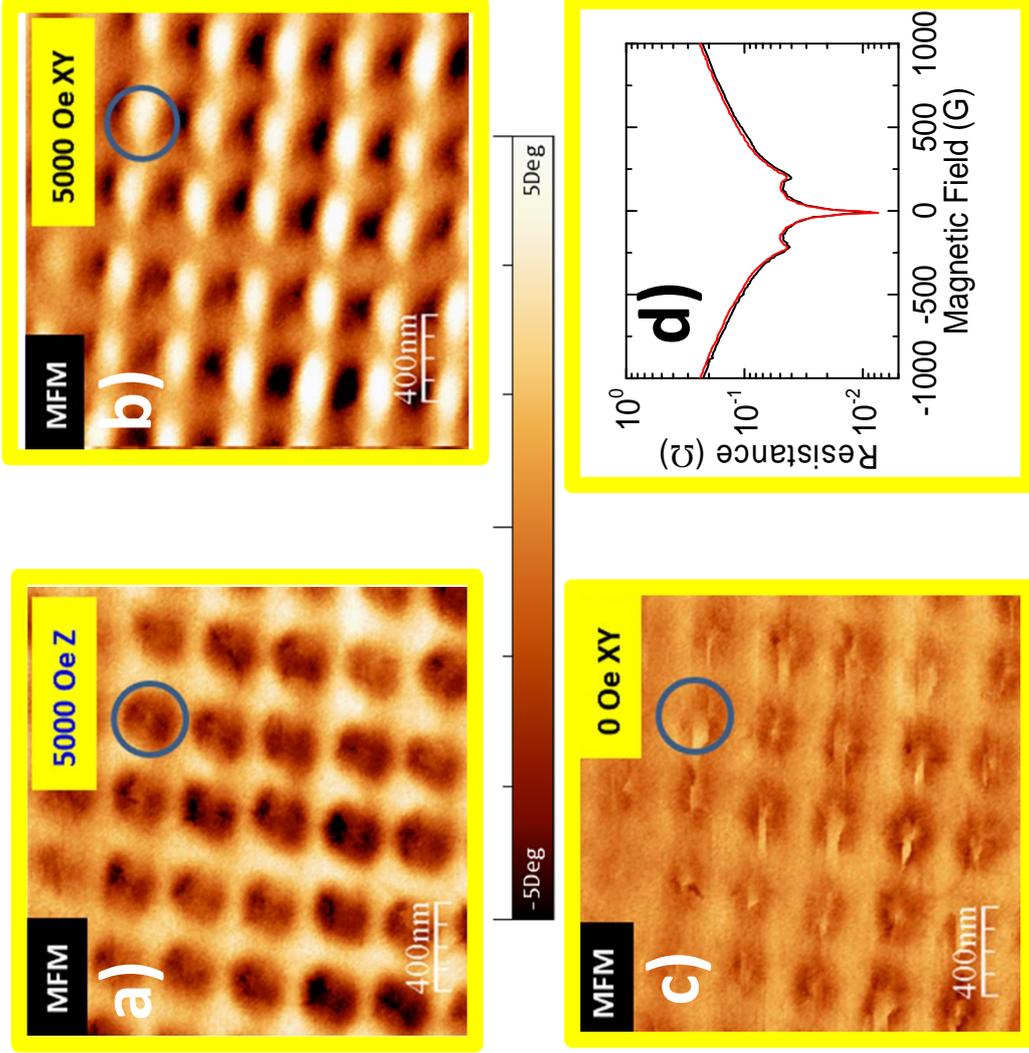

Figure 4

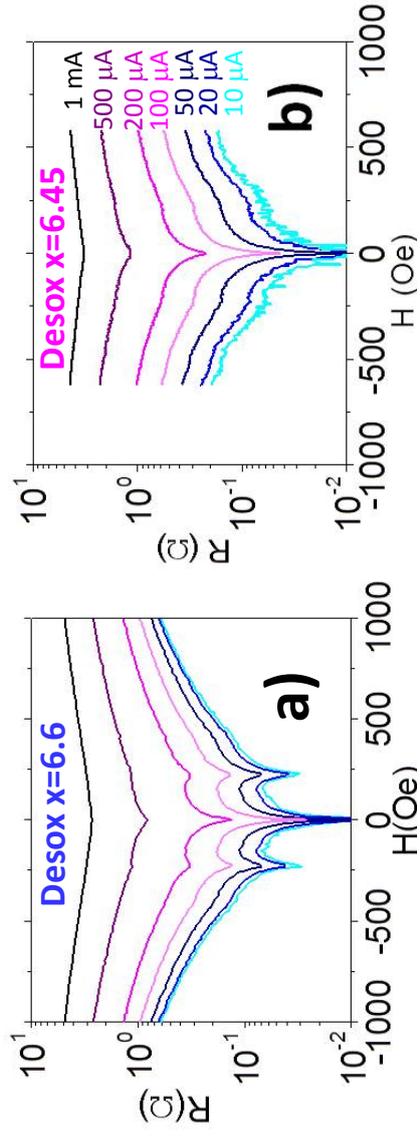
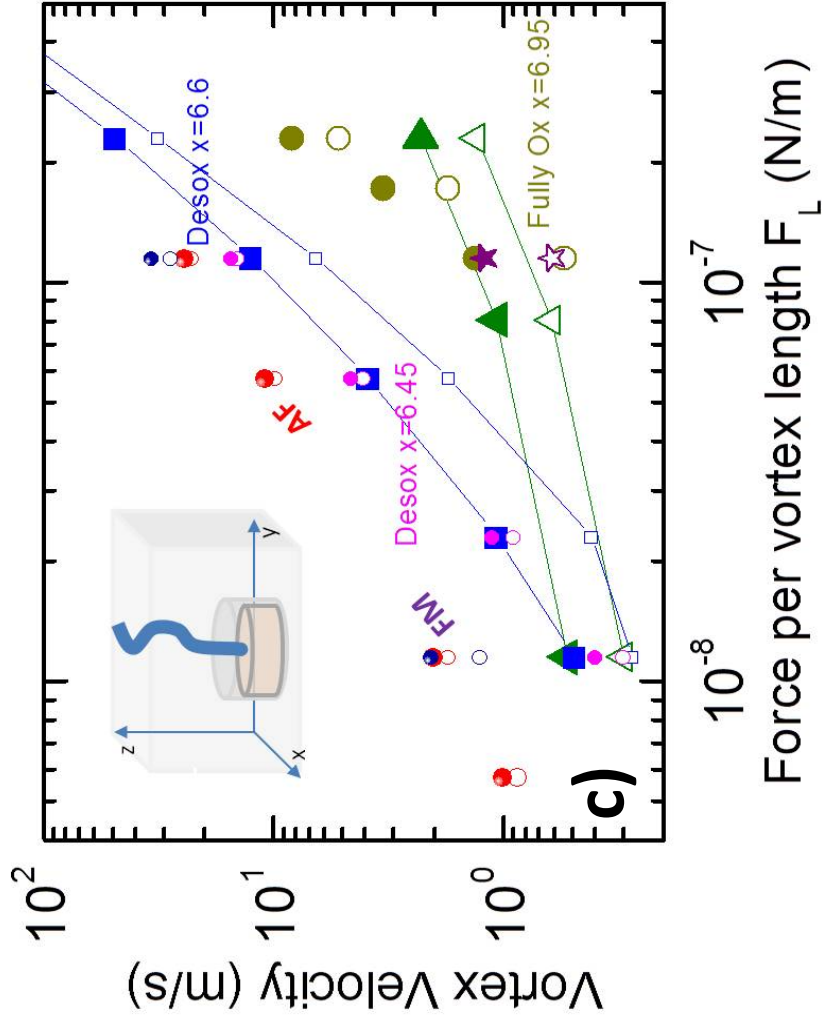